\documentclass[runningheads]{llncs}
\usepackage{graphicx}
\usepackage{amsmath}
\usepackage{amssymb}
\usepackage{float}
\usepackage[dvipsnames]{xcolor}

\begin{document}
	
	\title{Optimal risk in wealth exchange models: agent dynamics from a microscopic perspective}
	
	\titlerunning{Optimal risk in wealth exchange models}
	
	\author{Julian Neñer\inst{1} \and
		María Fabiana Laguna\inst{2}}
	
	\authorrunning{J. Neñer et al.}
	
	\institute{Instituto Balseiro, Universidad Nacional de Cuyo, R8402AGP Bariloche, Argentina	\email{julian.nener@ib.edu.ar}\\
		\and
		Centro At\'{o}mico Bariloche and CONICET, R8402AGP Bariloche, Argentina\\ \email{lagunaf@cab.cnea.gov.ar}}
	\maketitle              
	\begin{abstract}
		In this work we study the individual strategies carried out by agents undergoing transactions in wealth exchange models. We analyze the role of risk propensity in the behavior of the agents and find a critical risk, such that agents with risk above that value always end up losing everything when the system approaches equilibrium. Moreover, we find that the wealth of the agents is maximum for a range of risk values that depends on particular characteristics of the model, such as the social protection factor. Our findings allow to determine a region of parameters for which the strategies of the economic agents are successful.
		
	\end{abstract}
	\section{Introduction}
	
	It is a well known fact that different countries around the world exhibit highly unequal wealth distributions~\cite{uneven}. This phenomenon has not only been observed in many societies at different scales, but has been present repeatedly throughout history. 
	The relevance of this issue lies in the fact that economic inequality is the main cause of social tensions, with consequences in areas as diverse as health, crime, educational level or politics~\cite{Neckerman}.

	These observations lead to the following question: is this kind of behaviour inherent to societies themselves?
	
	It was only a century ago when the economist Vilfredo Pareto found repeating patterns describing power laws in the wealth distributions of different European countries~\cite{pareto}.The exponent of such distributions is nowadays referred as the Pareto index, and is a simple way to measure inequality. Later studies established that these power laws are only observed within the higher classes of the population, while the poorest individuals show log-normal or Gibbs distributions~\cite{gibbs}.
	
	With the goal of having a deeper understanding of the social behavior that leads to unequal wealth distributions, a wide variety of simple models based on ensembles of economic agents have been proposed for the past decades. These models are based on stochastic wealth flow processes between entities, where their time evolution is given by certain trading rules~\cite{book,macro1,kwem,kwp}.
	
	The macroscopic aspects concerning wealth exchange models have already been deeply studied. In particular, the Yard-Sale and Merger-Spinoff models, also referred to as the fair and loser rules respectively, have become quite popular, as they have shown promising results replicating empirical data~\cite{ysm1,ysm2,ysm3,ms,caon,benhur,risau,condens_ysm}. Nevertheless, as far as we know, the microscopic behavior hasn't been studied in the same detailed manner. A previous work done by Brian Hayes~\cite{hayes} looked into some of these aspects, finding that the first few transactions performed in the Yard-Sale model can determine the wealth positions of the agents for subsequent times, while on the Merger-Spinoff model a clear microscopic behavior could not be observed.
	The purpose of this paper is to address a  more in-depth study regarding the microscopic aspects of these multi-agent models, in order to analyze whether the success or failure of economic agents are determined by the model parameters.
	
	Previous theoretical work has been developed on some of these models by taking appropriate approximations and limits \cite{ysm3,theory1,theory2,theory3}, but in this manuscript we focus on a more empirical approach, as the increase in complexity due to the inclusion of a social protection factor and the non-linearities present in the evolution equations (as will be defined in the next section) makes it a very difficult task to develop analytical expressions for the models and goes out of the scope of this work.

	In the next section we describe the Yard-Sale and Merger-Spinoff models, along with the computational methods used to track the individual behavior of the agents. In Section \ref{sect:results} we present the obtained results with a brief description that will serve as a preview for Section \ref{sect:disc}, where we discuss on how this type of analysis leads to the finding of good and bad strategies in this kind of systems. 
	
	\section{Models}\label{sect:models}
	
	Consider a system of $N$ agents, each of them characterized by their wealth $w$ and risk propensity $r$. This last variable determines the fraction of wealth that they are willing to risk in every transaction, and is their only parameter of interaction. Its value remains constant throughout the wealth exchange process. It is also worth noting that the complement of this parameter, $\beta = 1 - r$, called risk aversion factor, appears more commonly in the literature. However, during the realization of this work the risk propensity $r$ (or simply risk) will be used for simplicity.
	
	Given a random pair $i$ and $j$, with wealths $w_i$ and $w_j$ respectively, their dynamics is described by
	
	\begin{equation}\label{eq:dynamics}
	\begin{matrix}
	w_i(t+1) = w_i(t) + (2\eta_{i,j}-1) \Delta w_{i,j} \\ 
	w_j(t+1) = w_j(t) - (2\eta_{i,j}-1) \Delta w_{i,j},
	\end{matrix}
	\end{equation}
	
	where $\eta_{i,j} \in \{0,1\}$ is a dichotomous random variable, and $\Delta w_{i,j}$ is a function defined by the model that depends on the risk and wealth of the transacting agents, as will become clear in the following subsections. It is worth noting that the total amount of wealth in the system is a conserved quantity.
	
	It has been proven that the only possible macroscopic equilibrium for this type of systems is the condensation of all wealth in a few agents~\cite{condens_ysm}. To alleviate this effect, an asymmetry is usually added to the distribution of $\eta_{i,j}$, that favors the poorest of the agents $i$ and $j$ in a single transaction:
	
	\begin{equation}
	p_{i,j} = \frac{1}{2} + f \dfrac{\left | w_i - w_j \right |}{w_i + w_j},
	\end{equation}
	
	where $f \in [0, \frac{1}{2}]$ has been called the social protection factor~\cite{f,f2}. The probability of the random variable $\eta_{i,j}$ of taking the value $k$, $\Pr(\eta_{i,j} = k)$, is then defined as:
	
	\begin{equation}\label{eq:etaij}
	\begin{aligned}
	\Pr(\eta_{i,j} = 0) = \Theta(w_i - w_j) p_{i,j} + \Theta(w_j - w_i)(1- p_{i,j})\\
	\Pr(\eta_{i,j} = 1) = \Theta(w_j - w_i) p_{i,j} + \Theta(w_i - w_j)(1- p_{i,j}),
	\end{aligned}
	\end{equation}
	
	where $\Theta$ is the heavyside step function.
	
	To measure how unequal is the wealth distribution of a certain society, the Gini index is commonly used. Given $N$ individuals, it is defined as the mean absolute difference between their wealths $w_i$:
	
	\begin{equation}\label{eq:ms}
	G = \dfrac{1}{2 N \sum_i w_i} \sum_{i = 1}^{N} \sum_{j = 1}^{N}\left | w_i - w_j \right |.
	\end{equation}
	
	This index establishes a clear measure of inequality, as $G \rightarrow 0$ when everyone has the same wealth, and $G \rightarrow 1$ when a single individual has the wealth of the whole system.
	
	In this work we present results for systems of $N = 10^4$ agents, with initial wealths and risks uniformly distributed in the interval $[0,1]$. The total amount of wealth is fixed at $W_\mathrm{tot} = 1$, and a minimum amount of wealth has been defined to allow for bankruptcy\footnote[1]{This value was chosen by taking into account the total amount of global wealth in USD, and the minimum currency division as 0.01 USD.} at $W_\mathrm{min} = 3 \times 10^{-17}$. We consider that an agent cannot recover once its wealth reaches a value less than $W_\mathrm{min}$, and consequently we set its wealth to zero. 
	
	A Monte Carlo step (MCS) is defined as $N/2$ transactions. The system evolves by iterating Eq. (\ref{eq:dynamics}) until a macroscopic equilibrium is reached. In general, unless specified otherwise, $10^3$ systems were evolved in parallel by using CUDA, a programming interface that allows high efficiency parallel computation in GPU~\cite{cuda}. 
	
	
	In what follows we present the functional form of $\Delta w_{i,j}$ for both models studied.
	
	\subsection{Merger-Spinoff model}
	
	Given a pair of agents $i$ and $j$ with wealths $w_i$ and $w_j$ and risks $r_i$ and $r_j$, a loser is chosen at random, who must give a fraction of its wealth
	to the opposing agent. Thus, the wealth transferred in every interaction will be
	
	\begin{equation}
	\Delta w_{i,j} = \eta_{i,j} r_j w_j + (1-\eta_{i,j}) r_i w_i.
	\end{equation}
	
	It is clear that in this model the agents lack the information of how much wealth is being risked by the opposing part, and as such, analogies have been made with marriages and divorces~\cite{hayes}, or thefts and frauds~\cite{ms}.
	
\begin{figure}[H]
		\centering
		\includegraphics[width=0.9\linewidth]{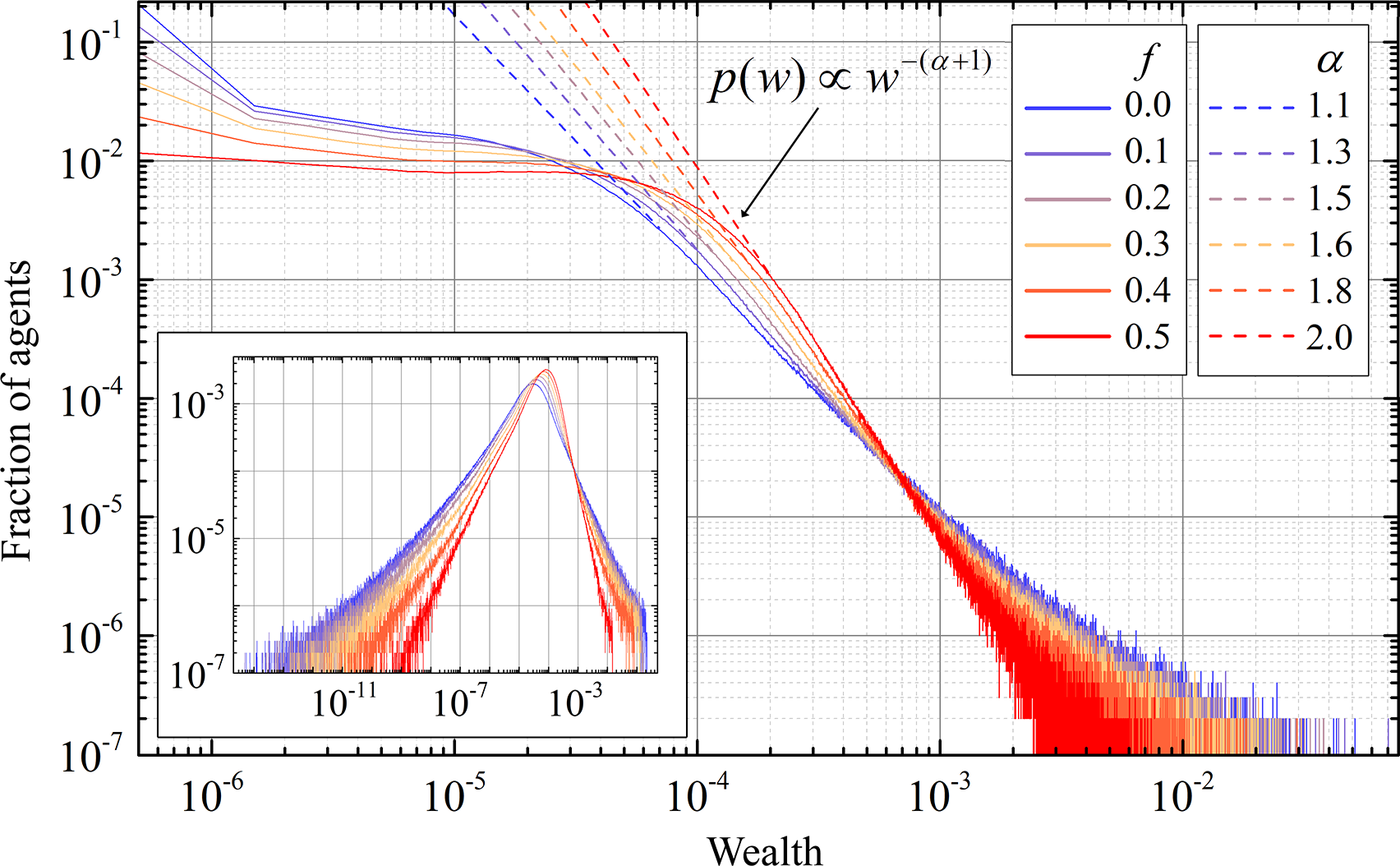}
		\caption{Merger-Spinoff model. Wealth distributions obtained averaging $10^3$ systems after equilibrium, for different social protection factors $f$. Power laws have been adjusted with their corresponding Pareto indices $\alpha$. The inset shows histograms with logarithmic binning.}
		\label{fig:pw_ms}
	\end{figure}

	To summarize the main macroscopic characteristics of this model, as previously studied in \cite{caon}, we present in Fig.~\ref{fig:pw_ms} the wealth distributions obtained in the stationary state for several values of the social protection factor $f$. Two different behaviors are observed: exponential curves for low wealth and power laws in the high-wealth region, with Pareto indices $\alpha$ between $1.1$ and $2$. The inset shows the same distributions taken with logarithmic binning, which allows to observe how the gap between the richer and poorer agents becomes progressively smaller with increasing $f$.

	\subsection{Yard-Sale model}
	
	In this model, initially proposed in~\cite{ysm0}, agents interact by risking a fraction of their wealth, as in the previous model. In this case, the amount of money exchanged corresponds to the minimum stake among the partners.
	Given a pair of agents with wealths $w_i$ and $w_j$, and risks $r_i$ and $r_j$, the amount of wealth transferred in a single transaction will be
	
	\begin{equation}
	\Delta w_{i,j} = \textup{min}(r_i w_i, r_j w_j).
	\end{equation}
	
\begin{figure}[H]
		\centering
		\includegraphics[width=0.9\linewidth]{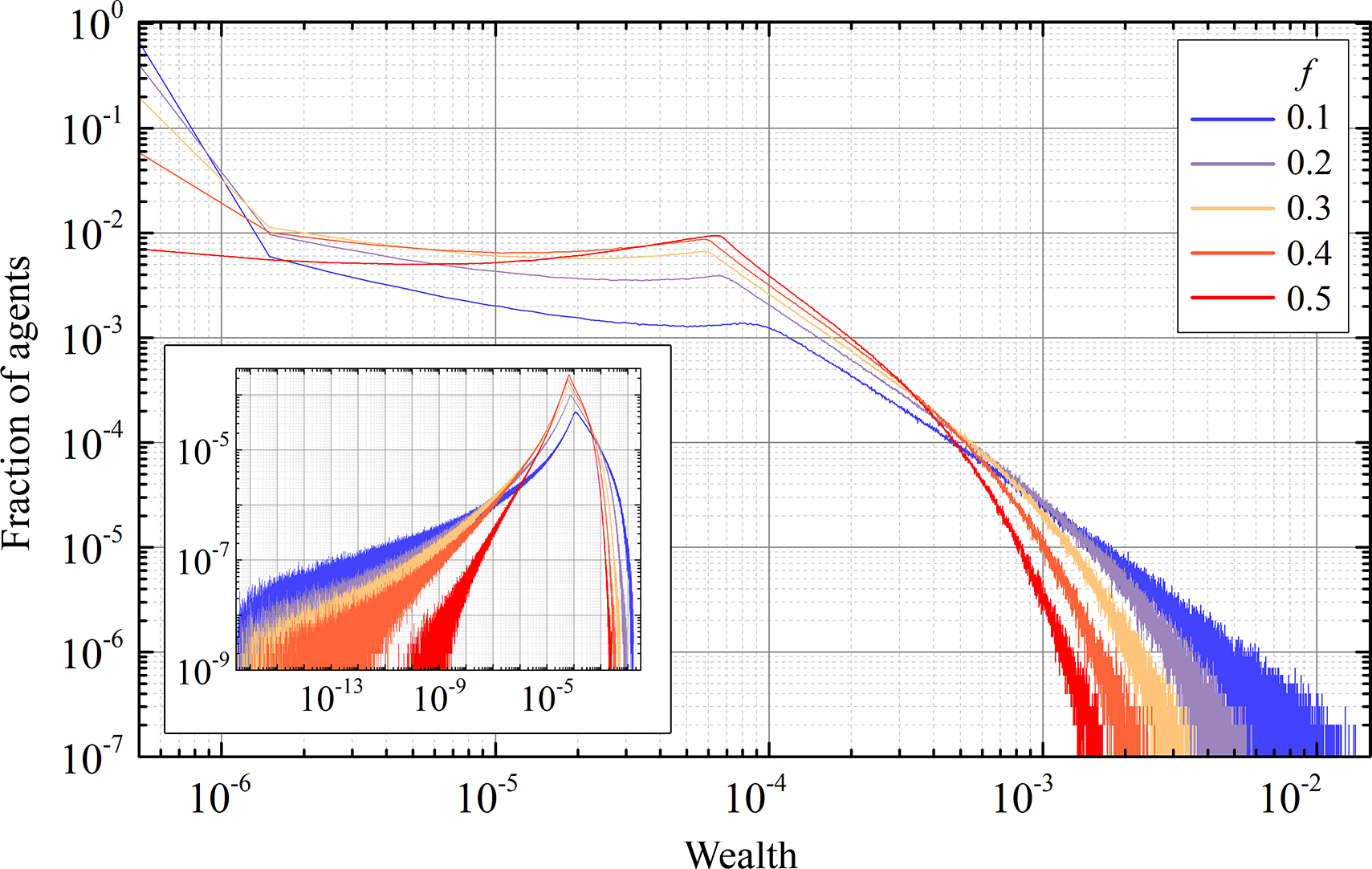}
		\caption{Yard-Sale model. Wealth distributions obtained averaging $10^3$ systems after equilibrium, for different social protection factors $f$. Exponential tails are observed. The inset shows histograms with logarithmic binning.}
		\label{fig:pw_ys}
	\end{figure}

	This model exhibits wealth distributions with exponential tails, and Pareto distributions can not be observed in any range of wealth, as shown in Fig.~\ref{fig:pw_ys}. This way, the model replicates the behavior of the low and middle classes, as it was stated in previous works \cite{benhur}. The inset shows the same distributions using logarithmic binning. The gap between the richer and poorer agents becomes smaller with increasing $f$ in a similar way as in the previous model.
	
	\section{Results}\label{sect:results}
	
	In what follows we present results regarding the microscopic aspects of the models described in Section \ref{sect:models}. 
	
	In all cases, the figures correspond to times taken after the wealth distributions had reached macroscopic equilibrium, where the equilibrium was defined as the time needed for the Gini index to become constant ($\sim 10^4$ MCS).
	
	\subsection{Merger-Spinoff model}
\begin{figure}[H]
		\centering
		\includegraphics[width=1\linewidth]{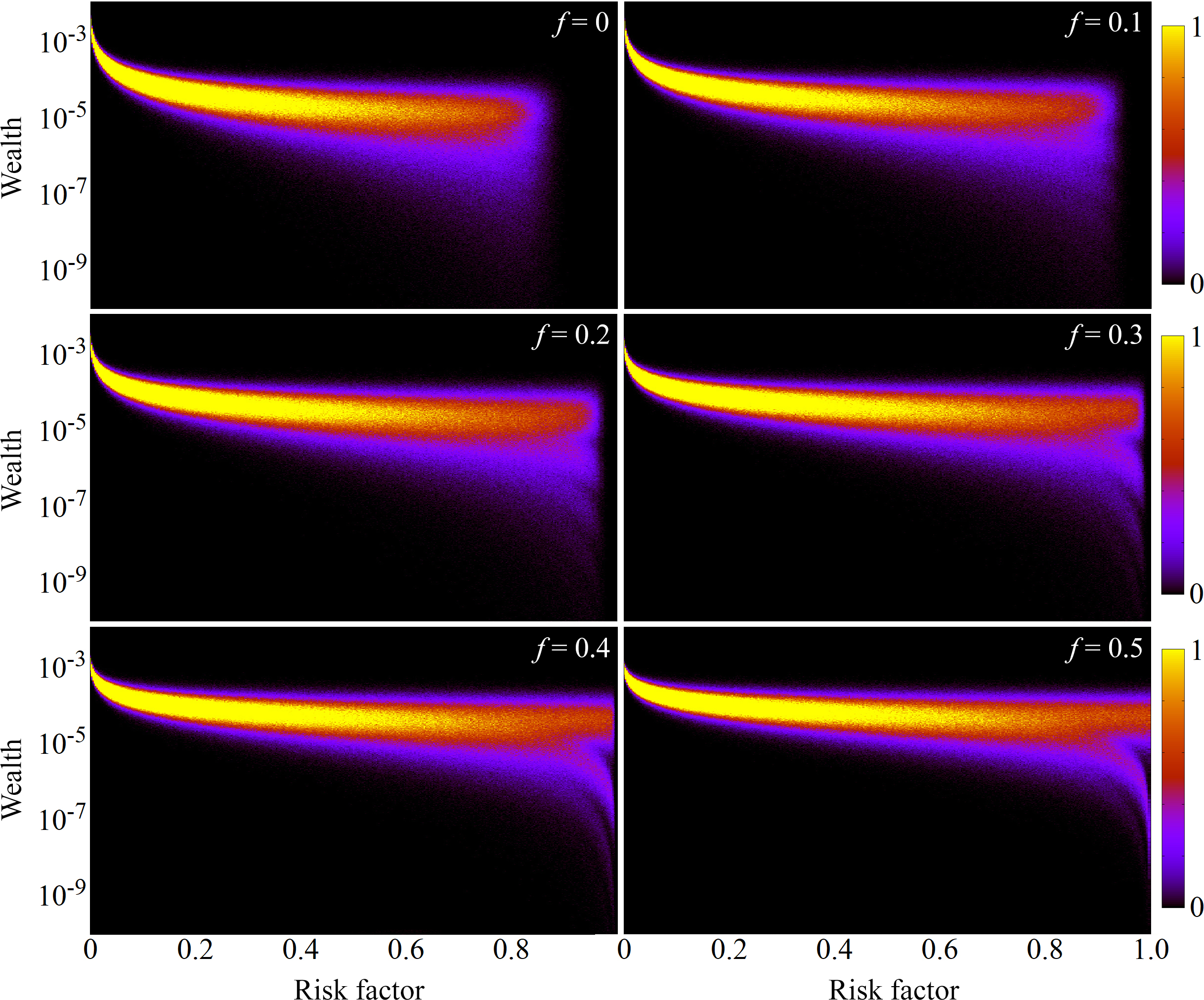}
		\caption{Density plots in the $r-w$ plane taking $10^3$ independent systems after equilibrium, for different social protection factors $f$. The color range indicates the relative concentration of agents in a given region.}
		\label{fig:heatmapslog_ms}
	\end{figure}

	With the goal of studying the individual behavior of the agents and how it leads to the well known macroscopic results, we record the final state of a large number of agents (obtained from $10^3$ simulations of $10^4$ agents each one). For a better visualization of the data, we define a density function $\rho(r,w)$ which takes increasing values with the amount of agents in a certain region of $r-w$ space. Each value of $\rho$ was then mapped to a given color. 
	In Fig.~\ref{fig:heatmapslog_ms} we show the density plots in $r-w$ space for different values of the social protection factor $f$.

	The density plots clearly show the existence of a a certain limiting value of $r$, such that those agents with a higher risk will always end up losing all their wealth before the system reaches equilibrium. We will refer to this value as $r = {\langle}r_\mathrm{crit}{\rangle}$, where the brackets refer to the spatial average made with all the systems; such averaged value is the one that can be inferred from the density plots.
	From Fig.~\ref{fig:heatmapslog_ms}, it is apparent that ${\langle}r_\mathrm{crit}{\rangle}$ increases with $f$, until it reaches its maximum possible value for $f>0.4$. This should be interpreted as follows: if the social protection factor is sufficiently high, even those agents with risks close to 1 will be able to keep some of their wealth. A consequence of this result is that higher values of $f$ allow for the observation of the evolution of higher $r$ agents, that would otherwise end up losing all their money in lower $f$ cases.
	
\begin{figure}[H]
		\centering
		\includegraphics[width=0.8\linewidth]{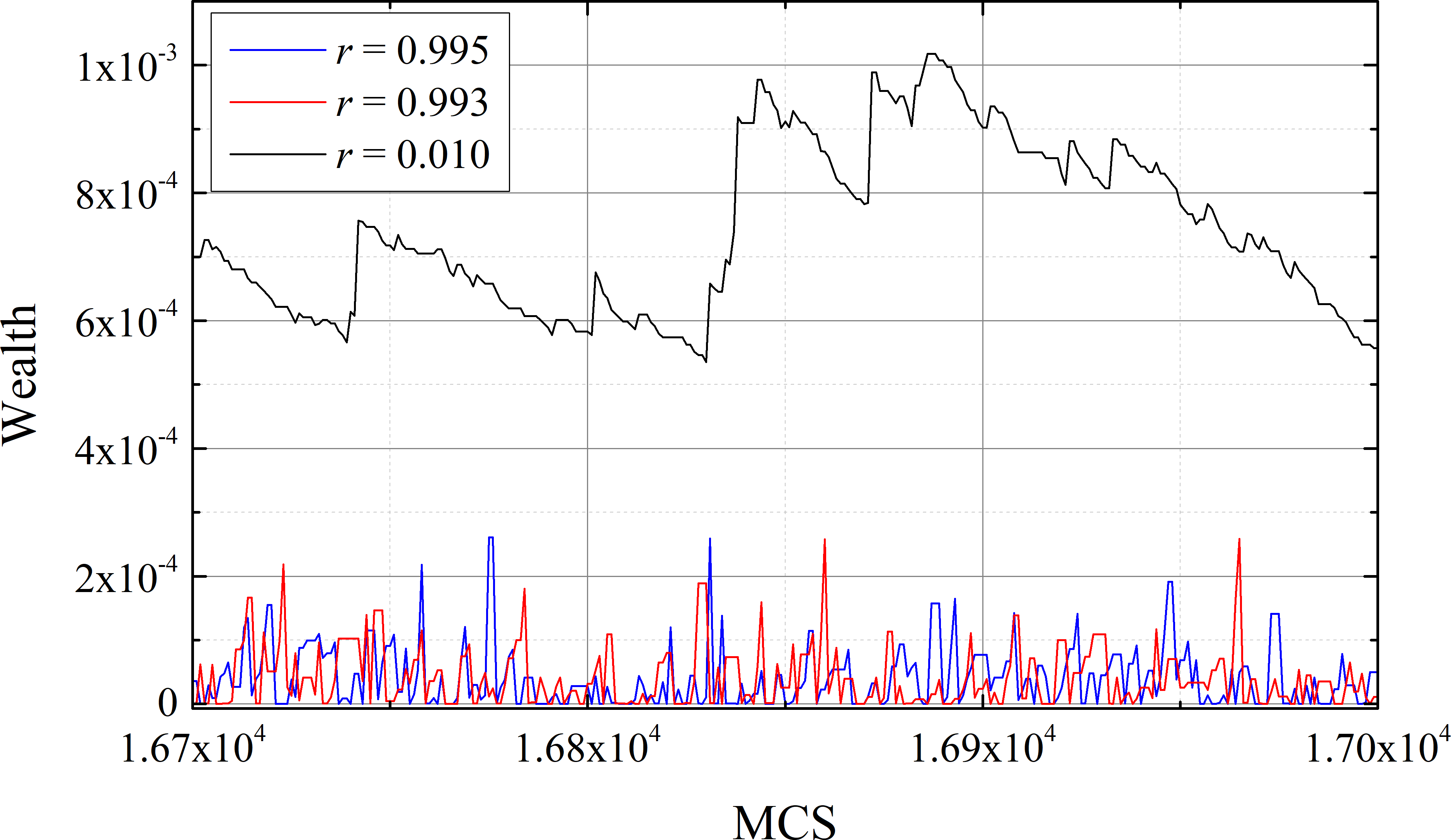}
		\caption{Comparison of the transaction history of two agents with high risk (red and blue), with an agent with very low risk (black). The two high-$r$ agents follow discrete dynamics, whereas the low-$r$ one presents a more continuous behavior.}
		\label{fig:highr_ms}
	\end{figure}

	
	Another interesting behavior can be observed in Fig.~\ref{fig:heatmapslog_ms} for the high risk region, where gaps in $w$ start appearing for increasing values of $f$. This can be understood by studying the transaction history of high-$r$ agents, as shown in Fig. \ref{fig:highr_ms}. Agents with a high value of $r$ follow a discrete dynamic in $w$ space, where alternating regions of high and low wealth are present depending on whether the transaction was successful or not. This is because high-$r$ agents have $r\sim 1$, making their risked wealth $rw$ comparable to their own wealth: $rw\sim w$, which excludes the possibility of obtaining intermediate wealth values. This behavior should be compared with the one of a low-$r$ agent, for which a more continuous\footnote[2]{We are not speaking of continuity in a rigorous way, but rather referring to the fact that the steps made in $w$-space are of a much smaller size.} dynamic is observed.
	

\begin{figure}[H]
		\centering
		\includegraphics[width=0.9\linewidth]{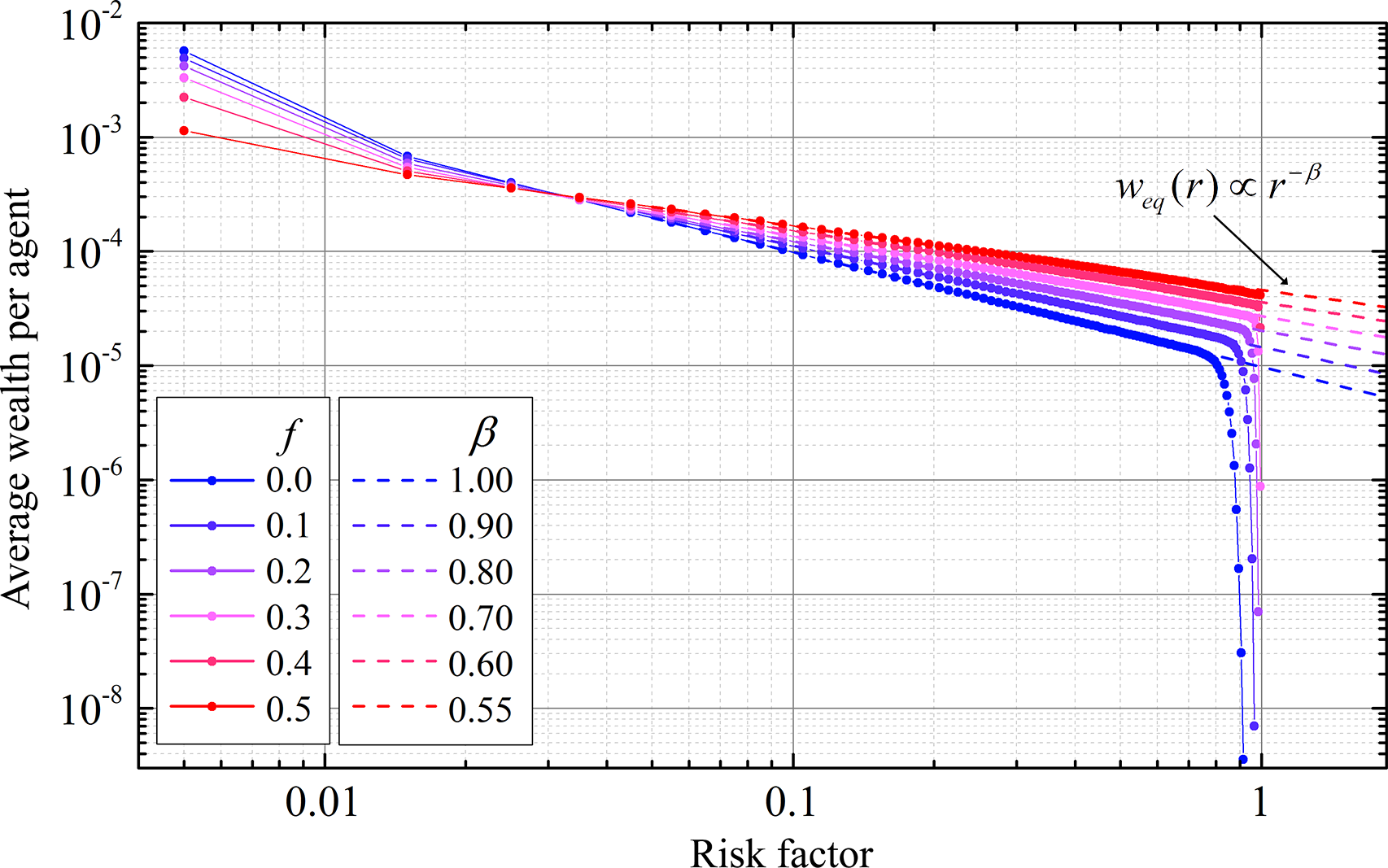}
		\caption{Curves of the average wealth per agent ${\langle}w_i{\rangle}$, as a function of the risk factor $r$, obtained averaging $10^3$ systems after equilibrium. Power laws were adjusted.}
		\label{fig:histor_ms}
	\end{figure}

	The heterogeneous distribution of individuals in the $w-r$ plane observed in the density plots of Fig.~\ref{fig:heatmapslog_ms} lead us to wonder if it is possible to find strategies that result in a higher individual wealth. In order to verify this assumption we plot in Fig.~\ref{fig:histor_ms} the histograms of the average wealth per agent as a function of their risk.
	
	These average wealth curves show a clear picture: there is a well defined function $w_\mathrm{eq}(r)$ that decreases monotonically with $r$. Power laws were adjusted of the form $w_\mathrm{eq}(r) = r^{-\beta}$. As expected, those agents with lower risk always end up with higher values of wealth.
	
	\subsection{Yard-Sale model}
	
	\subsubsection{Maxima in the Gini index.}
	
\begin{figure}[H]
		\centering
		\includegraphics[width=1\linewidth]{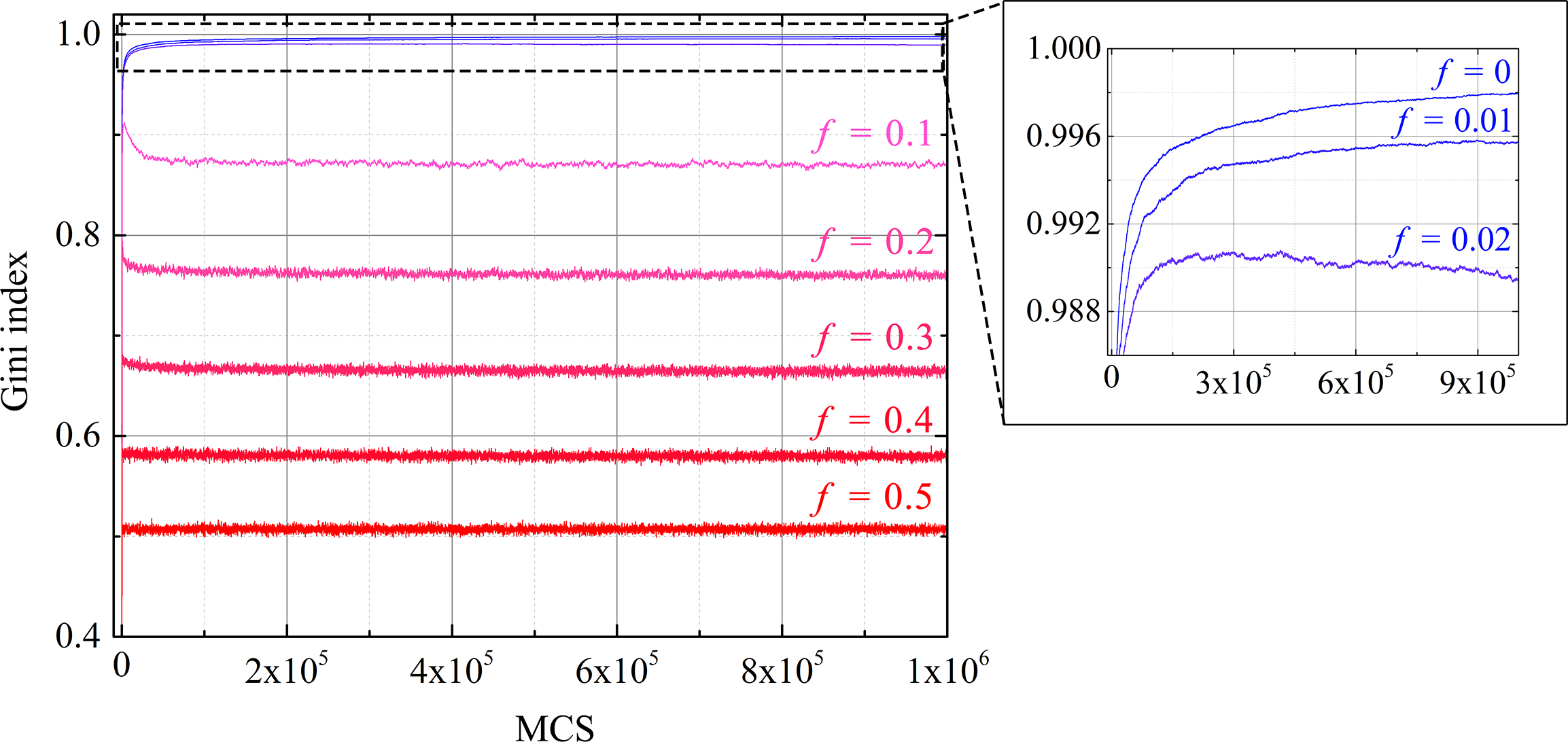}
		\caption{Gini index as a function of time for different values of the social protection factor $f$. A zoom of the curves has been made for small $f$ values.}
		\label{fig:gini_ys}
	\end{figure}

	An interesting feature of this model arises from observing the evolution of the Gini index during the wealth exchange process, as also seen in a previous work \cite{benhur}. We plot these curves in Fig.~\ref{fig:gini_ys} for different values of $f$ and note the presence of maxima for low values of $f$ within the first MCS. A zoom for small values of $f$ shows how these maxima become more prevalent in time as $f$ decreases.
	The systems converges at long times to lower Gini index values as $f$ increases, leading to less inequality. Note that the curve corresponding to the case $f = 0$ exhibits a positive derivative at all times and seems to converge to $1$, indicating that a single agent accumulates the wealth of the whole system.
	
\begin{figure}[H]
		\centering
		\includegraphics[width=0.9\linewidth]{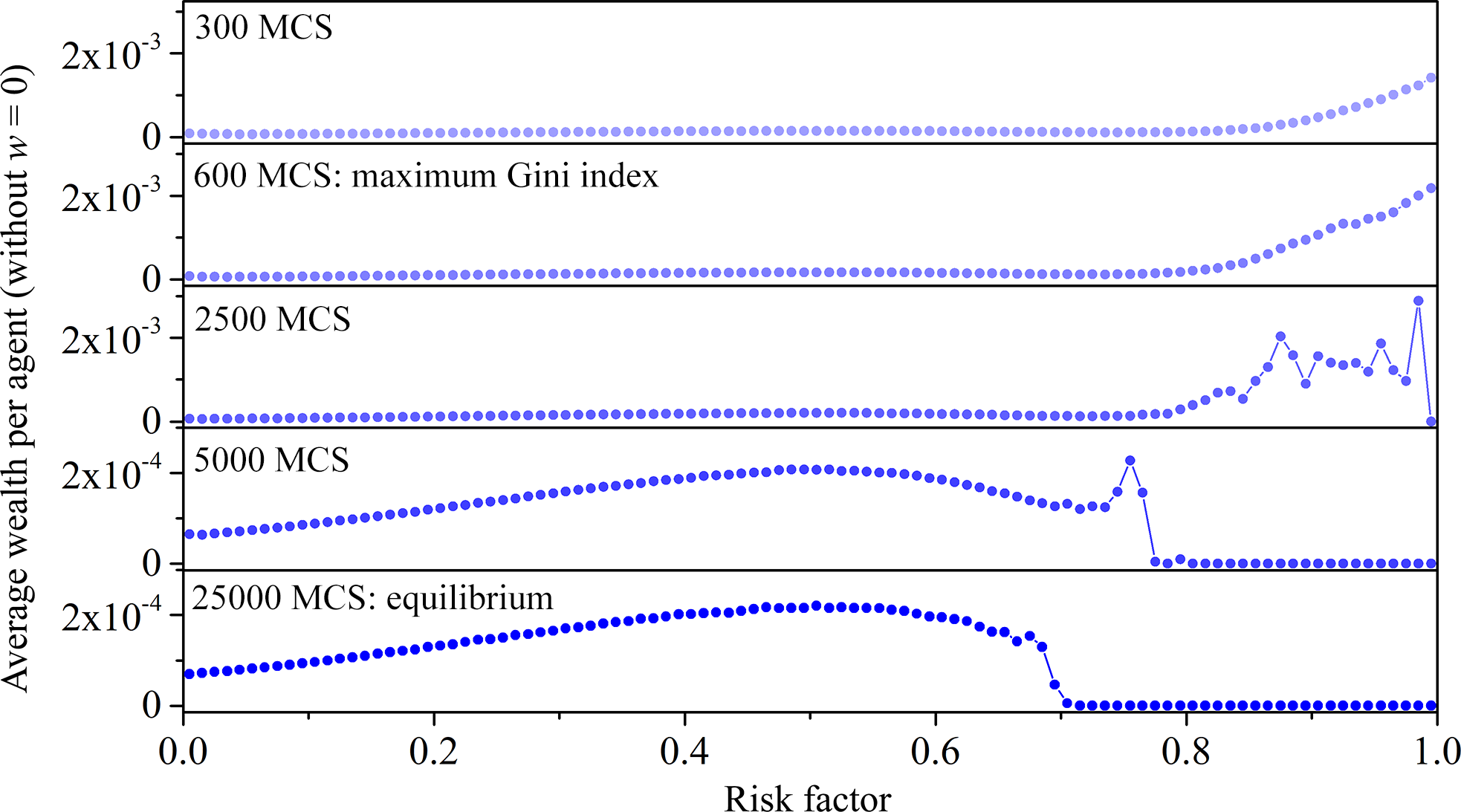}
		\caption{Time evolution of curves of the average wealth per agent, using $10^3$ systems, for a social protection factor of $f = 0.2$. A local maximum at $r \approx 0.5$ is present at all times (even if it is difficult to observe for shorter times). 
		}
		\label{fig:evtempgini_ys}
	\end{figure}

	To study the microscopic behavior that leads to the presence of the previously mentioned maxima in the Gini index, we plot in Fig.~\ref{fig:evtempgini_ys} the curves of the average wealth per agent at different times in the evolution of the system. In this figure, agents with $w=0$ were excluded. Moreover, we present results for a single value of the social protection factor, $f = 0.2$, but the same analysis applies to any other $f>0$. The times were chosen to compare the behavior of the system during the Gini index maximum, with the transient that is observed when the system evolves to equilibrium.
	
	The top panels of Fig.~\ref{fig:evtempgini_ys} show that during the first MCS those agents with higher $r$ accumulate more $w$. This is due to their possibility of earning more money in a single transaction. Then, it is precisely at the time corresponding to the maximum Gini index that this group accumulates the highest amount of wealth.
	
	When the system approaches equilibrium, these high-$r$ agents end up losing all their wealth, giving it up to the rest of the system. This behavior can be seen as the presence of a global maximum for $r = 1$ that becomes unstable. However, a local maximum located in a lower $r$ ($r \approx 0.5$) that is present for all times and thus stable, can also be observed. This value of $r$ is studied in more detail in the following subsections.
	
	\subsubsection{Correlations between risk and wealth.}
	
\begin{figure}[H]
		\centering
		\includegraphics[width=0.9\linewidth]{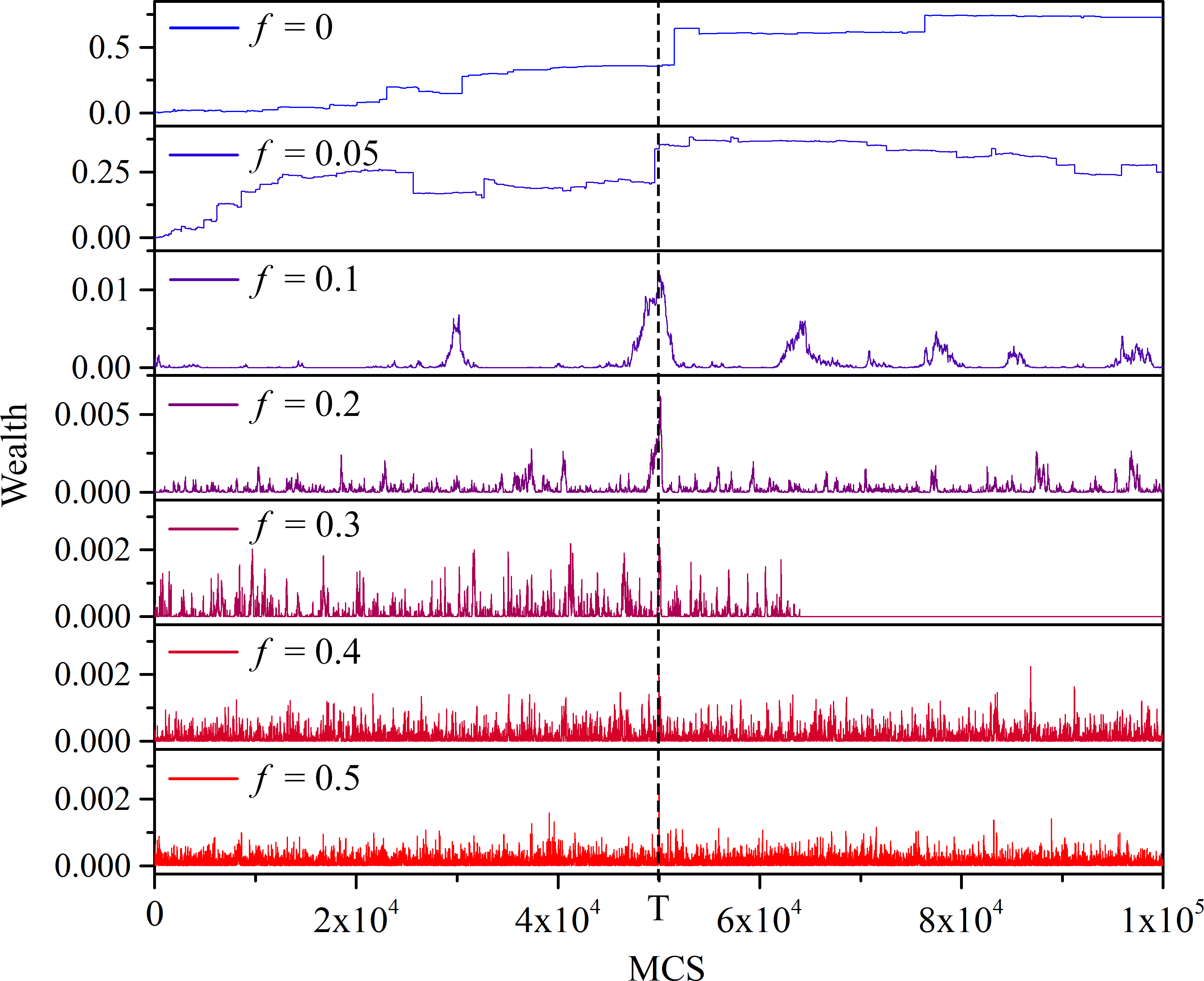}
		\caption{Transaction history of the richest agent at $T = 5 \times 10^4$ MCS. Each panel correspond to a system with a given social protection factor $f$, as indicated in the figure.}
		\label{fig:pathrichest_ys}
	\end{figure}

	Another possible indication of the existence of particular characteristics in the most successful agents can be found in the transaction history of the richest agent at a fixed time. We plot in Fig.~\ref{fig:pathrichest_ys} the transactions of the agent who was the richest at time $T = 5 \times 10^4$. This was done for systems with different values of $f$.
	
	These transaction histories show very different behaviors depending on $f$. For $f \leq 0.1$, periods of richness and poverty can be distinguished, but as $f$ approaches higher values, the fluctuations in $w$ space become too erratic and a distinction between poor and rich agents becomes unclear. This property explains why macroscopic equilibrium is reached faster with increasing $f$. 
	
\begin{figure}[H]
		\centering
		\includegraphics[width=0.9\linewidth]{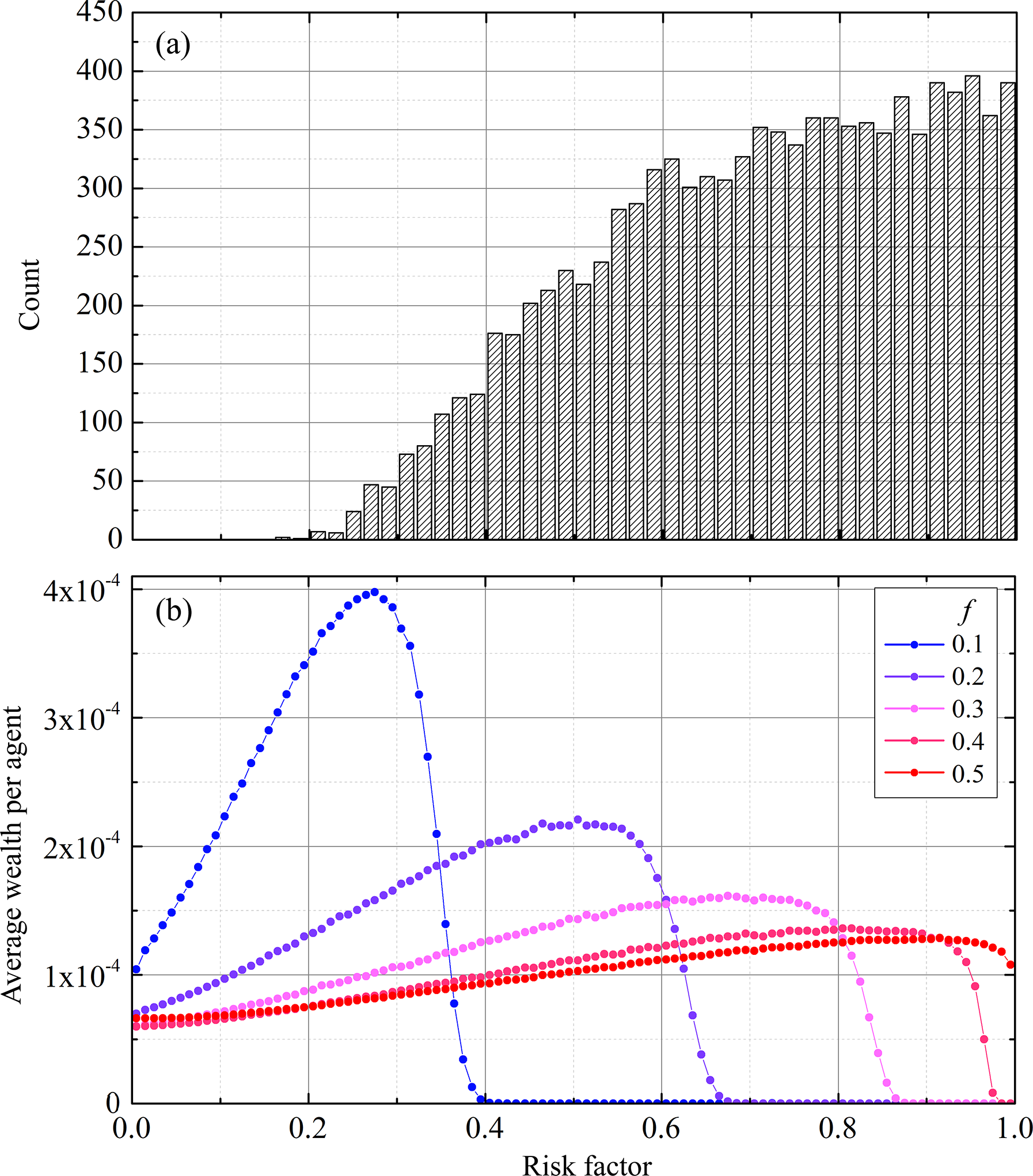}
		\caption{a) Risk histogram for the richest agent after $5 \times 10^4$ MCS, in $10^4$ systems with $f = 0$. b) Curves of the average wealth per agent, averaged over $10^3$ systems for several values of $f$ indicated in the plot.}
		\label{fig:riskhisto_ab}
	\end{figure}

	To verify the existence of any correlation regarding the risk of an agent and its possibility of success, we plot in Fig.~\ref{fig:riskhisto_ab} the results obtained for several values of $f$. The corresponding study was divided into two parts: for $f = 0$, since macroscopic equilibrium cannot be reached in a reasonable time scale, the study consisted in evolving $10^4$ independent systems and checking the risk of the richest agent after $5 \times 10^4$ MCS. The result is presented in Fig. \ref{fig:riskhisto_ab}-a. On the other hand, for $f > 0$ we plot in Fig.~\ref{fig:riskhisto_ab}-b the curves of the average wealth per agent after reaching equilibrium, as function of their risk.
	
	As we said, the final state of a system with $f = 0$ consists in a single agent that accumulates the wealth of the whole system. For this reason, the histogram shown in Fig.~\ref{fig:riskhisto_ab}-a, showing an increasing probability of an agent of becoming successful with increasing $r$, describes the existence of a good strategy for this system: it is convenient to have a high risk, such that a high enough wealth can be achieved in a relatively small amount of transactions. Once high enough $w$ is reached, interactions with the remaining agents will not affect the wealth of the richest agent in a significant manner.
	
	However, for $f > 0$, a maximum in the histogram can be observed for all $f$, indicating the existence of an optimal risk that allows an agent to obtain the maximum average wealth. It can also be noted that such a strategy will be less effective with increasing $f$.
	
	As it is apparent from the figure, an optimal strategy exists for all $f$, and it is radically different for $f = 0$ or $f > 0$, as the fluctuations in the wealth of the agents change the definition of success. 
	
	\subsubsection{The critical risk.}
	
	To investigate how the macroscopic results emerge from the microscopic behavior of the agents in the Yard-Sale model, density plots in the $r-w$ plane were made in a similar way as in section 3.1. 
	
\begin{figure}[H]
		\centering
		\includegraphics[width=1\linewidth]{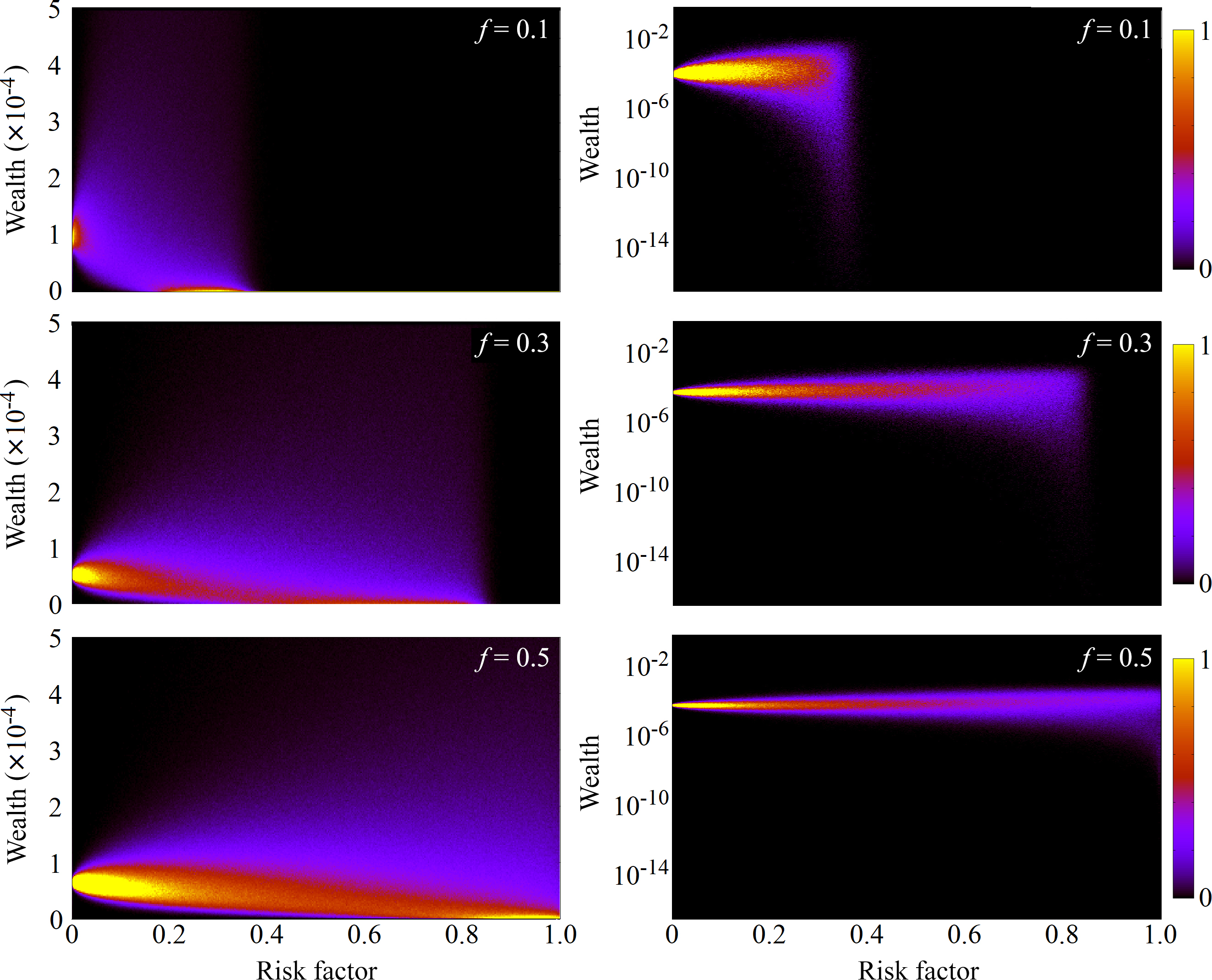}
		\caption{Density plots in the $r-w$ plane taking $10^3$ independent systems after equilibrium, for different social protection factors $f$. The color range indicates the relative concentration of agents in a certain region. Left panels: linear scale in the $w$ axis. Two regions of maximum density are observed. Right panels: logarithmic scale in the $w$ axis. The critical risk ${\langle}r_\mathrm{crit}{\rangle}$ is seen as the value such that the density is zero for all $w$ for any higher $r$.}
		\label{fig:heatmaps_ys}
	\end{figure}

	The density plots presented in Fig.~\ref{fig:heatmaps_ys} allow for a much more in-depth analysis of the behavior of the model. Choosing a linear (left panels) or logarithmic (right panels) representation of the $w$ axis allows for the study of different aspects of the system.
	
	Firstly, and in the same way as in the previous model, it is of special interest to note the existence of the critical risk factor ${\langle}r_\mathrm{crit}{\rangle}$, such that agents with $r$ above that value will always end up losing all their wealth when the system approaches equilibrium. The right panels show these values of $r$ very clearly.
	
	In the left panels, two regions of maximum density can be observed: the region near $r = 0$ shows that those agents with very low risk reach very similar values of $w$ when the system arrives at equilibrium. Such wealth values are close to the maxima previously observed in the wealth distributions of Fig.~\ref{fig:pw_ys}. On the other hand, the region of $r \lesssim {\langle}r_\mathrm{crit}{\rangle}$ shows agents with relatively high risks that still preserve some of their wealth, but didn't make good transactions. These results lead us to conclude that those agents with a low risk will always preserve a considerable amount of wealth, but they will never become rich. When agents become progressively more risky, a small number of them will be able to reach higher $w$ values, but most of the times they will end up losing almost everything.
	
	The previous results highlight the importance of studying the dependence of the critical risk ${\langle}r_\mathrm{crit}{\rangle}$ with the social protection factor $f$. For this purpose, we calculated the average time required for a specific agent to reach $w = 0$ by assigning it with increasing risk values, and averaging this result over $10^3$ independent runs. A time threshold had to be taken for each $f$, located after equilibrium. For $f > 0.1$, we choose a threshold of $5 \times 10^4$ MCS, while for $f < 0.1$ we take subsequent increasing times, as the time needed to approach equilibrium diverges when $f \rightarrow 0$, as previously shown in Fig.~\ref{fig:gini_ys}.
	
\begin{figure}[H]
		\centering
		\includegraphics[width=0.9\linewidth]{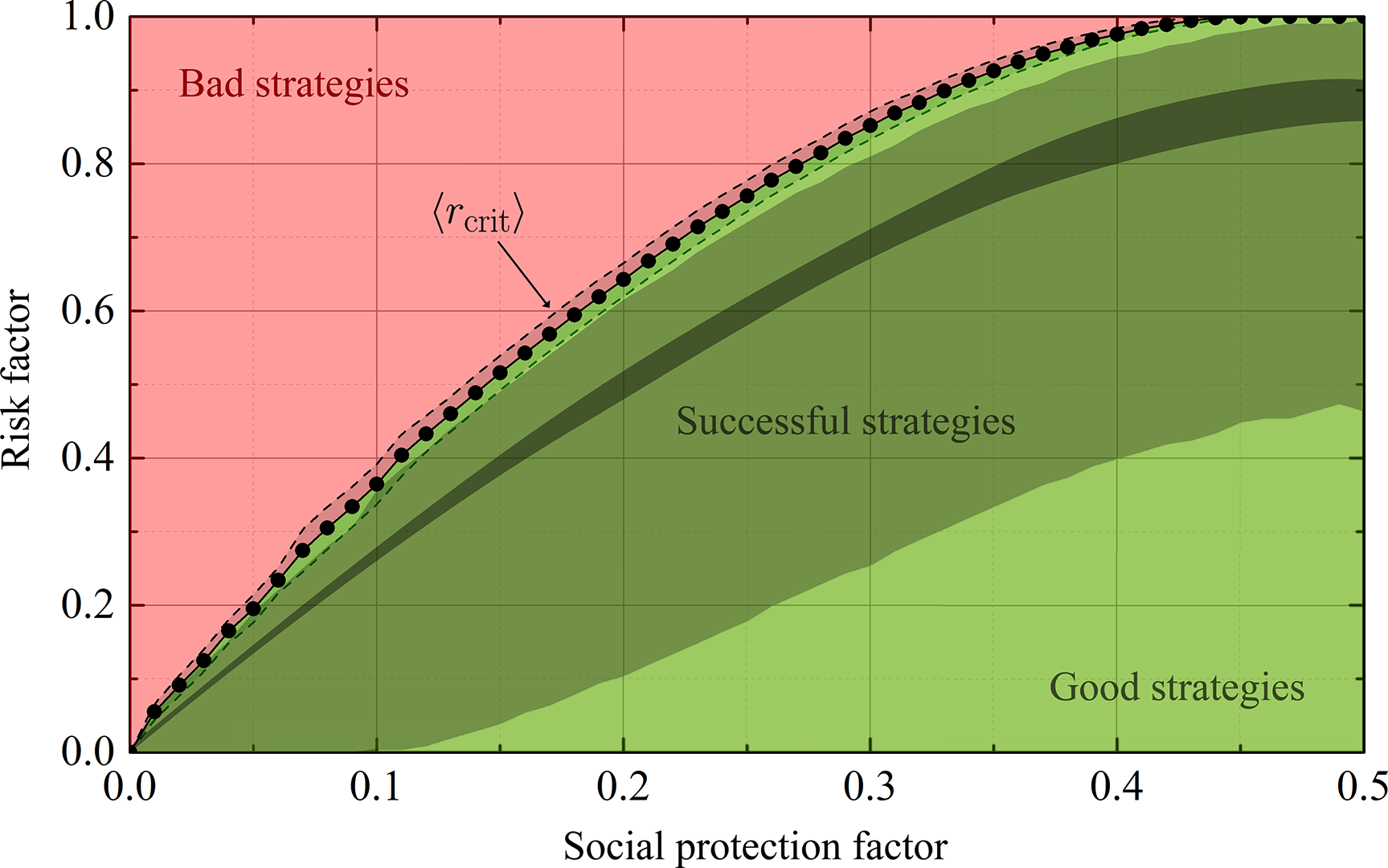}
		\caption{Average critical risk factor ${\langle}r_\mathrm{crit}{\rangle}$ as a function of the social protection factor $f$, representing the boundary between the good and bad strategies. Dotted lines show its standard deviation. The $r-f$ plane is divided in two main regions: the region of bad strategies, where agents reach $w = 0$ before equilibrium, and the region of good strategies, where agents never lose all their wealth. Successful strategies refer to agents that reach higher $w$ values than the average, and the region of darkest green depicts the optimal risk.}
		\label{fig:rcrit_ys}
	\end{figure}

	After repeating the previous process for different system sizes, a result of utter interest was obtained: the ${\langle}r_\mathrm{crit}{\rangle}$ curve, plotted in Fig.~\ref{fig:rcrit_ys}, turned out to be scale invariant, meaning that its shape does not depend on the amount of agents present in the system. The dashed lines indicate the standard deviation.
	
	
	As expected, the critical risk increases with $f$, indicating that the social protection factor allows for high-$r$ agents to stay with $w > 0$ even for long times after equilibrium. For values of $f \rightarrow 0.5$, ${\langle}r_\mathrm{crit}{\rangle}$ saturates to 1, indicating the inexistance of agents that lose all their wealth.

	
	The division observed in the $f-r$ plane shows a clear picture of what is considered a bad or a good strategy: the red region corresponds to those agents that reach $w = 0$ before the system arrives at equilibrium, and the green region corresponds to those agents that never lose all their wealth. Moreover, a subdivision in this last region was made depicting successful strategies, referring to those agents that reach higher wealth values than the average. Additionally, the optimal risk, corresponding to the wealth maxima previously observed in the curves of Fig. \ref{fig:riskhisto_ab}-b, is shown as the darkest green region. This region then indicates the set of parameters that will lead to the most successful strategy.
	
	Finally, since this last result does not depend on the size of the system, it can be concluded that this is an intrinsic property of the model.

	\section{Discussion}\label{sect:disc}
	In this work we studied microscopic aspects of two well known multi-agent systems: the so-called Yard-Sale and Merger-Spinoff models. We used computational methods to track individual agent behavior and found several interesting features that are hidden in the macroscopic studies of these models.

	\subsection*{Merger-Spinoff model}
	
	The density maps shown in Fig.~\ref{fig:heatmapslog_ms} unveiled the existence of gaps in wealth for those agents with high risk values. Higher values of the social protection factor $f$ allowed for the study of agents with higher risk propensity $r$, which for lower $f$ would end up losing all their wealth. This behavior could be explained by noting the discretized dynamics of the agents that risk all their wealth in every transaction, resulting in jumps in $w$ space of a size comparable to their own wealth.
	
	The density plots also hinted at the existence of a function $w_\mathrm{eq}(r)$ at equilibrium, which was then observed in the average wealth curves of Fig.~\ref{fig:histor_ms}. The monotonic decrease of $w_\mathrm{eq}(r)$ with risk indicates that agents with lower risk always end up with higher values of wealth. Moreover, the power laws obtained have an exponent that decreases with increasing $f$, reflecting increasingly egalitarian societies.
	
	The main conclusion for this model is that the most convenient strategy consists in risking the smallest fraction of wealth in every transaction. The reason becomes clear if we remember the definition of a transaction in this model, Eq. (\ref{eq:ms}): agents lose a fraction of their wealth that is proportional to their risk, and it is independent of the bet of the opposing agent. This makes those who risk less, lose less. 
	
	As a final comment on this model, it is important to mention that even though it allows to obtain good
    results at the macroscopic level, it clearly lacks interest when studied microscopically.
	
	\subsection*{Yard-Sale model}
	
	By plotting the time evolution of the Gini index, maxima before reaching equilibrium could be observed in the curves. The microscopic dynamics for times close to the maxima were then studied, as shown in Fig.~\ref{fig:evtempgini_ys}, which allowed to note how those agents with higher risks accumulate a high fraction of the total wealth in a relatively short time, but then end up losing everything, transferring their $w$ to lower risk agents.
	
	The transaction histories of the wealthiest agents for different $f$, as presented in Fig.~\ref{fig:pathrichest_ys}, showed how the social classes of the individuals can not be defined when $f$ is sufficiently high, and thus the position of the richest agent becomes irrelevant. However, a good strategy could be determined depending on $f$: when $f = 0$, the probability of an agent ending up hoarding the wealth of the whole system increases with its risk. For $f > 0$, an optimal risk dependent of $f$ could be found, such that the average wealth of the agents is maximized for all time.
	
	By studying the microscopic dynamics of the agents with density plots, as shown in Fig.~\ref{fig:heatmaps_ys}, a critical risk could be observed, such that those agents with higher $r$ than this value will always end up losing everything when the system reaches equilibrium. When this critical risk is plotted as a function of the social protection factor $f$, as it was presented in Fig.~\ref{fig:rcrit_ys}, a clear division in the $f-r$ plane allowed for a classification of good and bad strategies in this model, which turned out to be scale invariant. 
	The plot then shows how much wealth should be risked in different types of societies, characterized by their $f$, to always obtain the maximum possible profit.
	
	This last result has an interesting corollary: even in a society where the market regulation policies favor the poorest agents in every transaction as much as possible (which happens for $f = 0.5$), there will always be an individual optimal strategy that grants maximum average wealth. In other words, in such societies there are no agents ``out of the system" (i.e., with $w=0$) but there exist individuals who end up with more wealth than others, and they are those that have risks closer to the optimal one.
	
	Having found the existence of optimal strategies in this model, it remains as a subject of future work to study the behavior of agents that can change their risk factor at every time step. Different solutions could be found depending on the rational decisions made with the information they possess of their environment.
	
	\section*{Acknowledgements}
	The authors are grateful to M.N. Kuperman, G. Abramson and I. Sosa for their helpful suggestions and J.R. Iglesias for the careful reading of the manuscript and his valuable comments.
	
	This research did not receive any specific grant from funding agencies in the public, commercial, or not-for-profit sectors.
	
\end{document}